# Evidence-informed dynamic modelling and intervention prioritisation in diabetic bone marrow

Running title: Dynamic Modelling of Diabetic Bone Marrow

Paolo Madeddu, MD

Correspondence:
Professor Paolo Madeddu, MD
Bristol Medical School
University of Bristol
Bristol BS8 1TH
United Kingdom
Email: mdprm@bristol.ac.uk

## Abstract

**Aims/hypothesis.** Fragmented understanding of how diabetes disrupts interconnected bone marrow niches has hindered integrated strategies to restore function. Building on a previous static network, we developed a dynamic model to simulate intervention responses.

**Methods.** The investigator and ChatGPT dialogued to extract evidence from a frozen corpus and develop the model. A Python implementation extended Boolean logic to three ordinal levels (0, 1 and 2) representing biological states and, where applicable, intervention strength. Under high metabolic stress, 15 interventions were assessed at three strengths and four with binary settings; 171 pairs were tested at maximal strength. A domain-balanced score (0–100) integrated favourable and adverse changes. Staged simulations assessed withdrawal; regulatory constraints on recovery after metabolic stress reduction were examined separately.

**Results.** Endothelial reactive oxygen species (ROS) suppression achieved the highest score among 19 single interventions (24.2), improving three functional domains. Endothelial protein kinase B (Akt) augmentation scored 15.2 (two domains); oncostatin M (OSM) blockade and haematopoietic 66-kDa SHC-transforming protein 1 (p66Shc) loss each scored 13.6 (two domains). ROS suppression combined with either OSM blockade or p66Shc loss scored 37.9, (five domains). No pair restored all functions under high stress. Benefits of ROS suppression, Akt augmentation and ROS–OSM blockade disappeared after withdrawal. Reducing metabolic stress led to variable scores (range 4.5–100), increasing mainly with restored endothelial function, followed by progenitor recovery and interruption of inflammatory and adipocyte–stromal signalling.

**Conclusions/interpretation.** These simulations offer a rule-based framework for investigating diabetic marrow recovery. Experimental studies must establish whether it can guide targeted therapy.

## Introduction

Diabetes compromises endogenous repair, in part by impairing the mobilisation of reparative cells from the bone marrow (BM) into the circulation. Experimental and clinical evidence suggests that altered niche-dependent cell retention, disrupted endothelial barrier function, impaired endothelial-pericyte cross-talk, and increased inflammatory myelopoiesis contribute to this reparative deficit [1, 2, 3, 4]

Despite these mechanistic advances, how the marrow responds when one or more of its interconnected niches are restored remains poorly understood. A previous study reconstructed a static network of diabetic marrow regulation [5]. Such a network maps documented relationships; however, connectivity and node-deletion analyses alone cannot establish whether restoring one signal will recover physiological functions. Moving to dynamic simulations, Boolean models represent each component as active or inactive and apply logical rules to determine how activating and inhibitory signals change its state. Repeated application of these rules allows discrimination of how the effects of an intervention propagate through the network. Multi-valued logical models extend this framework beyond two states.[6] Here, each modelled activity or function can take a low, intermediate, or high level, allowing the model to distinguish partial improvement from complete recovery. The model can also explore whether improvements persist after an intervention is withdrawn, although it cannot predict how quickly they develop or how long they last. Explicit state definitions, regulatory rules and initial conditions make these predictions traceable to the supporting evidence and modelling assumptions.

The diabetic marrow literature provides several reasons to expect non-uniform recovery. Endothelial oxidative stress impairs marrow barrier integrity and angiocrine support, while pericyte dysfunction further compromises the vascular niche.[1, 2] Macrophage-derived oncostatin M (OSM) links inflammatory activity to stromal retention signalling.[7, 8] Adipocyte–stromal feedback can alter the supportive niche, while epigenetic changes in marrow progenitors can sustain an enhanced inflammatory response.[9, 10] Neural mechanisms add further regulation beyond a single chemokine pathway.[11, 12] These observations motivate a systems-level analysis using explicit regulatory rules to model interactions among marrow compartments and alternative formulations to explore uncertainty about the persistence and reversibility of dysfunction.

The present study extends the static network reconstruction through exploratory dynamic modelling developed through iterative dialogue between the investigator and an artificial intelligence (AI) assistant, informed by full-text evidence assessment. A three-level logical model represented functional responses, while alternative regulatory rules captured unresolved biological mechanisms. We first examined graded responses to 19 individual interventions under persistent high metabolic stress, followed by combinations targeting complementary pathways. Staged simulations then assessed whether improvements persisted after intervention withdrawal. Finally, we investigated the variable recovery following metabolic stress reduction, identifying regulatory conditions that favoured or constrained restoration. Together, these analyses aimed to prioritise candidate interventions, assess the persistence of their effects, and identify conditions that permit broader recovery of diabetic marrow function.

# Methods

## Study design and evidence extraction

The study followed a sequence from literature-based evidence extraction and dynamic model construction to intervention screening, assessment of functional recovery, intervention withdrawal and metabolic stress reduction (**Figure 1**).

The evidence base comprised 61 predefined corpus records, originally organised into 18 discovery and 43 validation records. Sixty publications were assessed in full text, and one abstract provided only contextual information. Evidence extraction recorded experimental comparisons, biological mechanisms, measured endpoints, interventions, timing and evidence of reversibility. Optical character recognition was used where necessary. Reviews, commentaries and the methods chapter informed interpretation but were not counted as independent replications of primary findings (**Supplementary Section S1**).

AI assistance was provided through iterative dialogue with ChatGPT (OpenAI; Astra, Extra High reasoning setting). ChatGPT assisted evidence extraction, proposed regulatory rules and alternative assumptions, and supported coding and simulation. The investigator defined the biological questions, reviewed the evidence in its experimental context, refined the rules and interpreted the results. Experimental findings were distinguished from modelling assumptions and transfers between biological contexts. Scientific decisions and final reporting remained the investigator's responsibility.

## Construction of the dynamic regulatory model

The original static network comprised 62 biological nodes and 79 directed interactions and was preserved as a separate reference. The dynamic model comprised 42 functional state variables governed by regulatory update rules and 20 external inputs or intervention controls. Variables were distinguished according to their cellular source and function. For example, endothelial C-X-C motif chemokine ligand 12 (CXCL12), OSM-induced stromal CXCL12 and C-X-C motif chemokine receptor 4 (CXCR4)-related retention were represented separately.

Experimental evidence was translated into logical rules describing how activating and inhibitory signals influenced biological functions. Jointly required inputs were combined using their minimum value (AND), alternative inputs using their maximum value (OR), and inhibitory relationships using inversion (NOT). The rules preserved experimentally observed differences between interventions. Endothelial protein kinase B (Akt) augmentation improved barrier integrity, directed mononuclear-cell passage and migration, but not spontaneous passage. Rho-associated coiled-coil-containing protein kinase (ROCK) inhibition restored fibroblast growth factor 2 (FGF2) but not Jagged expression, whereas reactive oxygen species (ROS) suppression restored both angiocrine outputs.[1] The requirement for the 66-kDa SHC-transforming protein 1 isoform (p66Shc) and signal transducer and activator of transcription 3 (STAT3) in OSM-induced stromal CXCL12 expression informed the corresponding signalling dependencies.[8] Persistent progenitor dysfunction after normal-glucose culture supported a persistence rule, while uncertainty about reversibility justified an alternative recovery rule.[15]

The model combined a common regulatory framework with 12 binary choices representing unresolved mechanisms. Their combinations generated $2^{12}$ (4,096) model variants, differing in assumptions about persistence or reversal of six injury states, autonomous OSM and adipocyte feedback, selected endothelial Akt responses, transfer of progenitor injury to marrow endothelium and the response to intermediate glucose stress. All variants shared the remaining regulatory rules and were evaluated without assigning probabilities to the alternative assumptions (**Supplementary Section S2**).

**Intervention grading and functional responses**

A common diabetic fixed point was established as the initial condition under high metabolic stress. Nineteen targeted interventions were selected. Intervention strength was classified as absent (0), intermediate (1) or maximal (2), where applicable. Six interventions used existing graded controls, while nine were extended to intermediate strength through exploratory perturbations of molecular targets or regulatory signals.

Four interventions—angiopoietin-2 blockade, Jumonji domain-containing protein 3 (Jmjd3) inhibition, acid sphingomyelinase inhibition and CXCR4 blockade—retained binary settings (absence or maximal intervention). Introducing intermediate settings for these four would have directly imposed changes on scored functional endpoints rather than allowing the regulatory network to generate the responses.

Comparative intervention-strength labels were distinct from the encoded control values. Maximal intervention could involve augmentation of an input to level 2, activation of a binary blockade at level 1, or complete target suppression to level 0. Intermediate perturbations were idealised modelling assumptions rather than calibrated doses or equivalent biological potencies. Their exact implementation is documented in **Supplementary Section S5**.

**Biological responses were represented independently of intervention strength.** Each functional variable assumed one of three ordinal states: 0 (low or absent), 1 (intermediate) or 2 (high). Higher values could indicate either improved function or greater dysfunction, depending on the variable. These states were not concentrations or comparable biological effect sizes.

All 19 individual interventions were examined at their permitted strengths, whereas all 171 unordered pairs were evaluated with both components at maximal strength. High metabolic stress was maintained throughout these intervention comparisons, allowing their effects to be assessed against an otherwise unchanged diabetic background. Metabolic stress reduction was examined separately in the final analysis.

**Simulation, scoring and computational checks**

A custom Python implementation using NumPy simulated the regulatory responses. At each synchronous update, every variable moved one ordinal level towards the target determined by its regulatory rule:

$$x_i(t + 1) = x_i(t) + \mathrm{sign}[F_i(x(t), u) - x_i(t)],$$

where $x_i$ denotes the current state, $F_i$ the regulatory target and u the external inputs. Simulations continued until a fixed point was reached, with a maximum of 160 iterations. Logical updates did not represent biological time.

Functional effects were assessed using 18 endpoints grouped into 11 domains: vascular integrity, endothelial migration, angiocrine support, pericyte support, myeloid inflammatory drive, excess OSM-linked retention, adipocyte-associated restraint, neural response competence, reparative-cell motility, stimulated inflammatory output and progenitor paracrine function. For each endpoint, the direction-adjusted change relative to the diabetic baseline was divided by two, corresponding to the full ordinal range. Endpoint contributions were averaged within each domain, and the 11 domain averages were weighted equally. The resulting index was multiplied by 100. This domain-balanced improvement score integrated the breadth and magnitude of favourable and adverse changes but did not represent the percentage of physiological function restored.

Interventions were ranked by their minimum improvement score across the 4,096 model variants, identifying effects maintained under the least favourable regulatory assumptions. The number of consistently improved domains was used to distinguish ties where applicable. For the 12-pair display, combinations tied at the cutoff were selected secondarily by their maximum scores. Sensitivity analyses examined 13 alternative weighting specifications (**Supplementary Section S5**).

Sixteen construction checks assessed whether the implementation reproduced the predefined regulatory relationships. Computational consistency was further examined using three asynchronous update schedules (seeds 11, 29 and 47). The original intervention-screen audit comprised 147,456 additional evaluations, excluding the exploratory intermediate-strength extension. No inferential tests were applied to deterministic outputs, and model frequencies were not interpreted as patient probabilities (**Supplementary Section S4**).

**Intervention withdrawal**

Staged simulations examined whether improvements persisted after withdrawal under continued high metabolic stress. The principal comparisons involved maximal endothelial Akt augmentation, ROS suppression and combined ROS suppression with OSM blockade.
Each intervention was maintained until a fixed point, then withdrawn without changing the metabolic background. The subsequent state was compared with the diabetic baseline.

Additional protocols examined temporary OSM and C-C chemokine receptor 2 (CCR2) blockade, alternative intervention sequences, withdrawal under reduced metabolic stress and renewed high stress. These analyses are documented in **Supplementary Section S6**.

**Metabolic stress reduction and recovery constraints**

Metabolic stress reduction was examined independently of targeted interventions. Starting from the common diabetic baseline, glucose stress was reduced from its initial high state (2) to intermediate (1) or low (0). Responses were assessed using the same functional score and alternative regulatory variants.

Three representative configurations illustrated progressively greater recovery: persistence of all six injury states with autonomous OSM and adipocyte feedback; reversal of endothelial injury alone; and reversal of all six injury states without autonomous feedback. The contribution of individual assumptions was assessed by comparing regulatory variants differing in only one choice, yielding 2,048 matched pairs per assumption. Differences in functional responses quantified their influence on predicted recovery.

Finally, 396 candidate condition–readout pairs were evaluated to identify measurements capable of distinguishing alternative recovery mechanisms. An eight-step procedure selected complementary readouts that progressively separated otherwise indistinguishable model variants. The analysis assumed exact ordinal observations without incorporating assay error, cost or feasibility. Recovery constraints and measurement selection are documented in **Supplementary Section S6**.

## Results

### Construction checks confirmed implementation of the regulatory rules

All 16 construction checks passed, confirming that the model reproduced the specified regulatory relationships. These included distinct endothelial responses to Akt, ROCK and ROS perturbations, separation of endothelial and pericyte Akt effects, and p66Shc- and STAT3-dependent stromal CXCL12 induction. Additional checks reproduced the expected responses to RAMP1 loss, forskolin bypass, CCR2 blockade and EGFR loss, together with stability of the diabetic baseline (**Supplementary Section S4**).

The original synchronous simulations reached fixed points, and all tested asynchronous schedules reproduced the corresponding endpoints. Agreement was maintained across the intervention-screen and staged-withdrawal audits. These findings supported computational consistency without establishing independent biological validity.

### Individual interventions produced graded and selective functional improvements

Under persistent high metabolic stress, endothelial ROS suppression achieved the highest minimum improvement score among the 19 individual interventions, increasing from 12.1/100 at intermediate strength to 24.2 at maximal strength. Maximal suppression consistently improved three functional domains: vascular integrity, endothelial migration and angiocrine support. Angiocrine recovery remained partial because endothelial CXCL12 was not restored (**Figure 2**).

Endothelial Akt augmentation ranked second, with scores of 7.6–10.6 at intermediate strength and 15.2–21.2 at maximal strength. It improved barrier integrity, directed mononuclear-cell passage and endothelial migration, while spontaneous cell passage remained elevated. Additional improvement in FGF2 and Jagged depended on the alternative regulatory assumptions. OSM blockade and haematopoietic p66Shc loss each scored 6.8 at intermediate strength and 13.6 at maximal strength, consistently improving myeloid inflammatory drive and excessive OSM-linked retention. ROCK inhibition increased its score from 6.1 to 12.1, improving two domains.

Other interventions produced more restricted responses. Maximal pericyte Akt augmentation and angiopoietin-2 blockade each scored 9.1, improving one domain. Suppression of S100A8/A9, RAGE or NF-κB yielded scores of 4.5–18.2, with broader benefits depending on regulatory assumptions. Some intermediate perturbations produced no consistent improvement, while four interventions retained binary responses (**Figure 2; Supplementary Section S5**).

Overall, greater intervention strength generally increased functional improvement, but responses remained selective. Endothelial redox suppression produced the broadest consistent benefit, whereas other interventions improved specific vascular, inflammatory, stromal or reparative functions. No individual intervention restored all assessed functions under persistent high metabolic stress.

### Paired interventions extended recovery to complementary functional domains

Screening of 171 intervention pairs identified two leading combinations: endothelial ROS suppression with either OSM blockade or haematopoietic p66Shc loss. Both achieved a minimum improvement score of 37.9/100 and consistently improved five of the 11 functional domains, compared with three domains after ROS suppression alone (**Figure 3A**).

Seven additional combinations of ROS suppression with pericyte Akt augmentation, angiopoietin-2 blockade, stromal p66Shc loss, STAT3 blockade, Jmjd3 blockade, acid sphingomyelinase blockade or CXCR4 blockade shared the next-highest score (**33.3**), each improving four domains. Combinations with S100A8/A9, RAGE or NF-κB suppression achieved a minimum score of 28.8 and reached 42.4 under favourable regulatory assumptions (**Supplementary Section S5**).

The two leading pairs showed identical functional responses. Both retained the improvements in vascular integrity, endothelial migration and angiocrine support associated with ROS suppression, while extending recovery to myeloid inflammatory drive and excessive OSM-linked retention (**Figure 3B**). Thus, the highest-ranking combinations brought together complementary vascular and inflammatory effects. However, no pair restored all assessed functions under persistent high stress.

**Intervention benefits disappeared after withdrawal under persistent high metabolic stress**
We next examined whether selected intervention benefits persisted after withdrawal. Endothelial ROS suppression, Akt augmentation and combined ROS suppression–OSM blockade produced improvement scores of 24.2, 15.2–21.2 and 37.9, respectively, while active. Following withdrawal, all regulatory variants returned to the constructed diabetic baseline, with an improvement score of zero (**Figure 4**). The benefits of these interventions therefore depended on continued treatment in the modelled high-stress environment. Additional staged protocols are reported in Supplementary Section S6.

**Metabolic stress reduction revealed variable recovery determined by regulatory conditions**
Finally, we examined recovery following metabolic stress reduction alone. Decreasing glucose stress from level 2 to 1 generated 240 distinct endpoint configurations. S100A8/A9, RAGE and NF-κB consistently declined to intermediate levels, while colony-stimulating-factor activity and other functional responses remained dependent on regulatory assumptions.

Further reduction to level 0 generated 320 distinct configurations. The glucose-dependent inflammatory pathway was consistently suppressed, whereas recovery of endothelial, progenitor, pericyte, neural and other marrow functions remained variable. Consequently, maximal metabolic stress reduction produced improvement scores ranging from 4.5 to 100 (**Figure 5A**). Three representative regulatory conditions illustrate this variability.

When established dysfunction persisted across all six injury states, and OSM, and adipocyte feedback remained self-sustaining, recovery was limited to a score of 4.5. Allowing endothelial dysfunction alone to resolve increased the score to 31.8, despite persistence of the remaining abnormalities. Complete restoration of the 18 assessed endpoints (score 100) occurred only when all six injury states could resolve, and neither feedback circuit remained autonomous.

Matched-rule comparisons confirmed the predominant influence of endothelial cell recovery (**Figure 5B**). Allowing endothelial dysfunction to resolve increased the mean improvement score by 21.2 points, followed by progenitor recovery (15.2). Removal of self-sustaining OSM or adipocyte feedback increased the score by 13.6 and 9.1 points, respectively.

The analysis also identified eight complementary readouts that distinguished alternative recovery mechanisms and persistent dysfunction. Together, they separated the 4,096 regulatory variants into 256 response signatures, leaving 16 variants per signature. The measurements covered endothelial CXCL12, pericyte support, OSM-linked myeloid activity, adipocyte-associated restraint, adrenergic responses, reparative-cell motility, stimulated interleukin-12 and progenitor paracrine function. Across all 396 candidate condition–readout pairs, 2,560 signatures could be distinguished, although some regulatory variants remained observationally equivalent (**Supplementary Figure S2; Supplementary Section S6**).

## Discussion

This study extends the structural reconstruction of diabetic BM regulation into a dynamic model for examining functional responses to intervention. The simulations indicate that targeting individual regulatory pathways can improve selected marrow functions without restoring the entire niche.

Suppressing endothelial ROS produced the highest minimum improvement score among the individual interventions, while combining it with OSM blockade or haematopoietic p66Shc loss extended recovery in inflammatory and retention-related functions. These improvements required continued intervention under persistent high metabolic stress. In contrast, reducing metabolic stress may permit broader recovery, though its extent depends strongly on the reversibility of established dysfunction and the persistence of self-sustaining regulatory signals. Together, these findings distinguish the mechanisms governing selective functional rescue from those determining wider marrow recovery.

**Single interventions reveal distinct opportunities and limits to recovery**
The graded intervention analysis shows why targeting a single signalling pathway within a highly connected cellular compartment does not necessarily restore all its functions. Endothelial barrier integrity, directed mononuclear-cell passage, spontaneous passage and angiocrine support arise from overlapping but distinguishable mechanisms.[1]

Endothelial ROS suppression produced broader vascular improvement, encompassing barrier function, migration and angiocrine support. This intervention maintained the highest minimum improvement score across the alternative regulatory rules and remained the leading intervention under different weighting specifications. Likewise, Akt augmentation improved barrier integrity, directed passage and endothelial migration without correcting excessive spontaneous passage. The response of FGF2 and Jagged remained dependent on alternative regulatory assumptions. ROCK inhibition showed a more restricted pattern, consistent with experimental evidence distinguishing its effects on FGF2 from those of ROS suppression on both FGF2 and Jagged.[1] However, endothelial CXCL12 support remained impaired despite maximal redox suppression. Thus, even the highest-ranking intervention did not restore every function of the targeted compartment.These differences emphasise that improvement in one endothelial function may coexist with persistent defects in others.

The distinction between endothelial and pericyte regulation further illustrates the limits of individual interventions. Pericyte dysfunction was not corrected by endothelial Akt augmentation, reflecting the separate cellular origins and regulatory mechanisms underlying vascular support.[2] From a systems biology perspective, marrow cell populations respond to shared metabolic stresses but retain distinct functional dependencies. The vascular-niche literature similarly distinguishes endothelial survival, angiocrine signalling and the maintenance of functional haematopoietic stem cells.[13] Improvement in one component therefore cannot be assumed to restore the broader niche.

**Complementary interventions broaden functional recovery**
The paired-intervention analysis suggests that broader recovery may require simultaneous modulation of distinct regulatory pathways. ROS suppression primarily improved vascular function, whereas OSM blockade reduced inflammatory amplification and excessive progenitor retention. Their combination therefore extended recovery beyond the endothelial compartment. The comparable improvement obtained by combining ROS suppression with either OSM blockade or haematopoietic p66Shc loss suggests that distinct interventions can influence overlapping inflammatory and retention-related processes.

These mechanisms show how different marrow populations can contribute to persistent dysfunction. Macrophage-derived OSM promotes stromal CXCL12 signalling and can reinforce CXCR4-dependent progenitor retention.[7, 8] Haematopoietic p66Shc contributes to the inflammatory signalling associated with this response, while stromal p66Shc and STAT3 participate in OSM-dependent CXCL12 induction.[8] These processes are related but should not be treated as a single pathway acting identically across cellular compartments.

The clinical translation of such combinations raises questions of target specificity and safety. In particular, OSM signalling may have physiological functions beyond the pathological responses represented here. A phase 2 study of an OSM-neutralising antibody in systemic sclerosis did not confirm clinical efficacy, and higher exposure was associated with reductions in haemoglobin and platelet counts.[20] These observations emphasise the need to establish whether suppression of pathological marrow signalling can be achieved without compromising normal haematopoiesis.

**Intervention withdrawal reveals dependence on continued support**

The withdrawal simulations distinguish improvement requiring continued intervention from durable resetting of the underlying regulatory system. Under persistent high metabolic stress, the benefits of endothelial Akt augmentation, ROS suppression and combined ROS–OSM blockade disappeared after intervention withdrawal. Thus, correcting selected signalling pathways did not eliminate the ongoing influence of the diabetic environment.

This finding suggests that functional rescue may remain dependent on sustained pathway modulation when the initiating metabolic stress persists. However, logical iterations do not represent biological time, and these simulations cannot establish treatment duration, the kinetics of recovery or whether prolonged exposure could induce additional adaptive changes.

**Metabolic stress reduction reveals the importance of injury reversibility**

Reducing metabolic stress provides a complementary perspective on recovery. Unlike targeted interventions, which modify selected pathways while the diabetic environment remains unchanged, metabolic improvement reduces an upstream stimulus affecting multiple marrow compartments. However, removing a damaging stimulus and reversing established injury are distinct biological processes.

An important finding was the discrepancy between consistent molecular responses and highly variable functional recovery. Reducing metabolic stress suppressed the glucose-dependent S100A8/A9–RAGE–NF-κB inflammatory pathway across the regulatory alternatives. Nevertheless, under the least favourable assumptions, the improvement score remained only 4.5, substantially below the minimum score produced by direct endothelial ROS suppression under persistent high stress. This difference should not be interpreted as evidence that metabolic correction is intrinsically less effective than targeted intervention. The scores describe different modelled processes rather than comparative therapeutic efficacy. Targeted ROS suppression reliably improved selected vascular functions despite continuing metabolic stress. By contrast, metabolic improvement affected the initiating stimulus but did not necessarily reverse dysfunction already established within the marrow.

When endothelial injury was allowed to resolve, recovery increased substantially. Additional improvement occurred when progenitor injury and other abnormalities were reversible and OSM or adipocyte feedback no longer sustained pathological signalling. Complete restoration of the assessed functions was possible only under the most favourable combination of these assumptions. Thus, the wide range of recovery after reducing metabolic stress reflected uncertainty about biological reversibility rather than differences in the intensity of metabolic correction.

The dominant influence of endothelial injury reversibility is particularly relevant to diabetic BM disease. Endothelial dysfunction may simultaneously compromise vascular integrity, cell trafficking, and angiocrine support, thereby influencing several aspects of the surrounding niche. Restoring endothelial function could therefore facilitate broader recovery, although the model does not show that this process occurs spontaneously after glucose normalisation. Persistent progenitor abnormalities represent a second constraint. Evidence that progenitor dysfunction can persist after cells are transferred to a normal-glucose environment, along with epigenetic alterations affecting inflammatory responses, supports investigating how prior metabolic exposure shapes subsequent recovery.[10, 15] These observations do not establish permanent damage but suggest that correcting the extracellular environment may be insufficient to reverse every acquired cellular abnormality.

Self-sustaining inflammatory and adipocyte–stromal signalling provides a related mechanism. OSM-mediated interactions may maintain excessive stromal retention signals, while adipocyte-associated feedback can reinforce changes in marrow composition and haematopoietic support.[8, 9] Once established, such interactions could continue to influence marrow function after the initiating metabolic stimulus declines. Their reversibility therefore requires investigation independently of glucose reduction.

The same consideration extends to other compartments. Acid sphingomyelinase inhibition has been shown to modify reparative-cell motility, but it remains uncertain whether metabolic correction can reverse the corresponding dysfunction.[16] Neural regulation introduces additional mechanisms involving sympathetic and sensory signalling that cannot be reduced to a single chemokine pathway.[11, 12, 17] Studies in physiological and ageing marrow provide relevant evidence, although recovery of these systems in diabetic BM remains incompletely characterised.

Taken together, these findings suggest that metabolic improvement and targeted interventions address different aspects of diabetic marrow dysfunction. Metabolic correction reduces the initiating stress, whereas targeted interventions may directly modify functional abnormalities that persist despite metabolic improvement. This distinction supports investigating complementary therapeutic strategies, including whether metabolic improvement affects the durability or extent of targeted repair. Such combinations should be tested experimentally rather than inferred to be beneficial from the present simulations alone.

**Complementary measurements could clarify incomplete recovery**

The measurement analysis identified observations that could distinguish alternative explanations for incomplete recovery after metabolic stress reduction. These included persistent endothelial dysfunction, insufficient pericyte support, continued inflammatory signalling, adipocyte-associated restraint, neural abnormalities and progenitor dysfunction.

Assessing several of these functions together could be more informative than relying on a single biomarker. For example, suppression of glucose-responsive inflammatory signalling might coexist with impaired endothelial CXCL12 support or persistent progenitor dysfunction. Similarly, reduced OSM-dependent stromal retention would not necessarily demonstrate recovery of endothelial angiocrine function. The model-selected eight complementary readouts provide a framework for investigating these possibilities. Several measurement combinations were equally informative, allowing future studies to consider experimental practicality when selecting assays.

## Limitations and mitigations measures

AI can transfer terminology across disciplines, introducing mathematical modelling terms into biological descriptions with potentially unintended implications. To reduce ambiguity, the investigator reviewed and replaced terms such as “relaxation”, “resetting”, “fixed point”, “autonomous feedback” and “pulse”, with words typical of molecular and cellular biology.

The model integrates findings from different species, disease settings and assays. Source-specific evidence records and investigator review helped distinguish observed relationships from extrapolations and modelling assumptions. Nevertheless, independent review would strengthen evidence interpretation, particularly when relationships are transferred between cellular contexts. A source-linked framework for this assessment is provided in **Supplementary Section S7**. Predictions also depend on the mechanisms represented, assumptions about recovery and a common diabetic starting state. The 4,096 alternative rule sets and matched-rule comparisons assessed sensitivity to specified uncertainties, but cannot address omitted mechanisms or establish the relative biological likelihood of the alternatives.

The common starting state supports consistent intervention comparisons but does not capture patient heterogeneity. Intervention rankings depend on the selected outcomes, their equal weighting across 11 domains and the greater experimental detail available for some pathways. Alternative weighting schemes were examined to assess ranking sensitivity, although this does not remove differences in evidence coverage.

The 0–100 score is an ordinal comparison measure, not a calibrated percentage of marrow recovery. Complete restoration of assessed functions at low metabolic stress was conditional on favourable assumptions about the reversibility of dysfunction and persistence of feedback.

Implementation checks and comparisons across alternative update schedules supported computational consistency but did not constitute independent biological validation. Testing covered only a finite set of schedules, and measurement selection assumed error-free observations. Further development should incorporate biological timing, treatment exposure, patient variability and measurement uncertainty. Safety assessment requires representation of drug toxicity, off-target effects and a broader range of deterioration from baseline. Independent experiments are therefore needed to evaluate the prioritised interventions, proposed measurements and predicted persistence of improvement after treatment withdrawal.

### Conclusions and perspectives

This study shows how investigator-guided, AI-assisted modelling can bring scattered biological findings into a common framework to prioritise interventions and design experiments. By linking predicted responses to explicit evidence and assumptions, the approach identifies both opportunities for recovery and uncertainties that most need experimental resolution.

The principal translational implication is that reducing the metabolic insult and reversing established marrow dysfunction may require complementary strategies. Future experiments should examine endothelial redox rescue, combined vascular and inflammatory interventions, and the conditions permitting recovery after metabolic improvement. These studies will determine whether the model-identified regulatory mechanisms can guide therapeutic approaches to restore the diabetic marrow microenvironment. (Figure 6).

**Acknowledgements**

The author acknowledges the assistance of ChatGPT (OpenAI) with evidence extraction, regulatory rule formulation, coding and simulation, as described in Methods. The author reviewed and interpreted all outputs and assumes full responsibility for the manuscript.

**Funding**

No external grant funding was received for this work.

**Declaration of competing interest**

The author declares no competing interests related to this work.

**Data and code availability**

The supporting regulatory model, numerical outputs and analysis code are supplied in the accompanying supplementary material and reproducibility archive.

**CRediT authorship contribution statement**

Paolo Madeddu: Conceptualization, Methodology, Investigation, Formal analysis, Validation, Writing – original draft, Writing – review and editing.

**Declaration of generative AI and AI-assisted technologies**

ChatGPT (OpenAI; Astra, Extra High reasoning setting) assisted with evidence extraction, rule formulation, code development and manuscript preparation. OpenAI image generation was used to produce the conceptual illustration in Figure 6. The author reviewed and edited all AI-assisted material and takes full responsibility for the accuracy and integrity of this work.

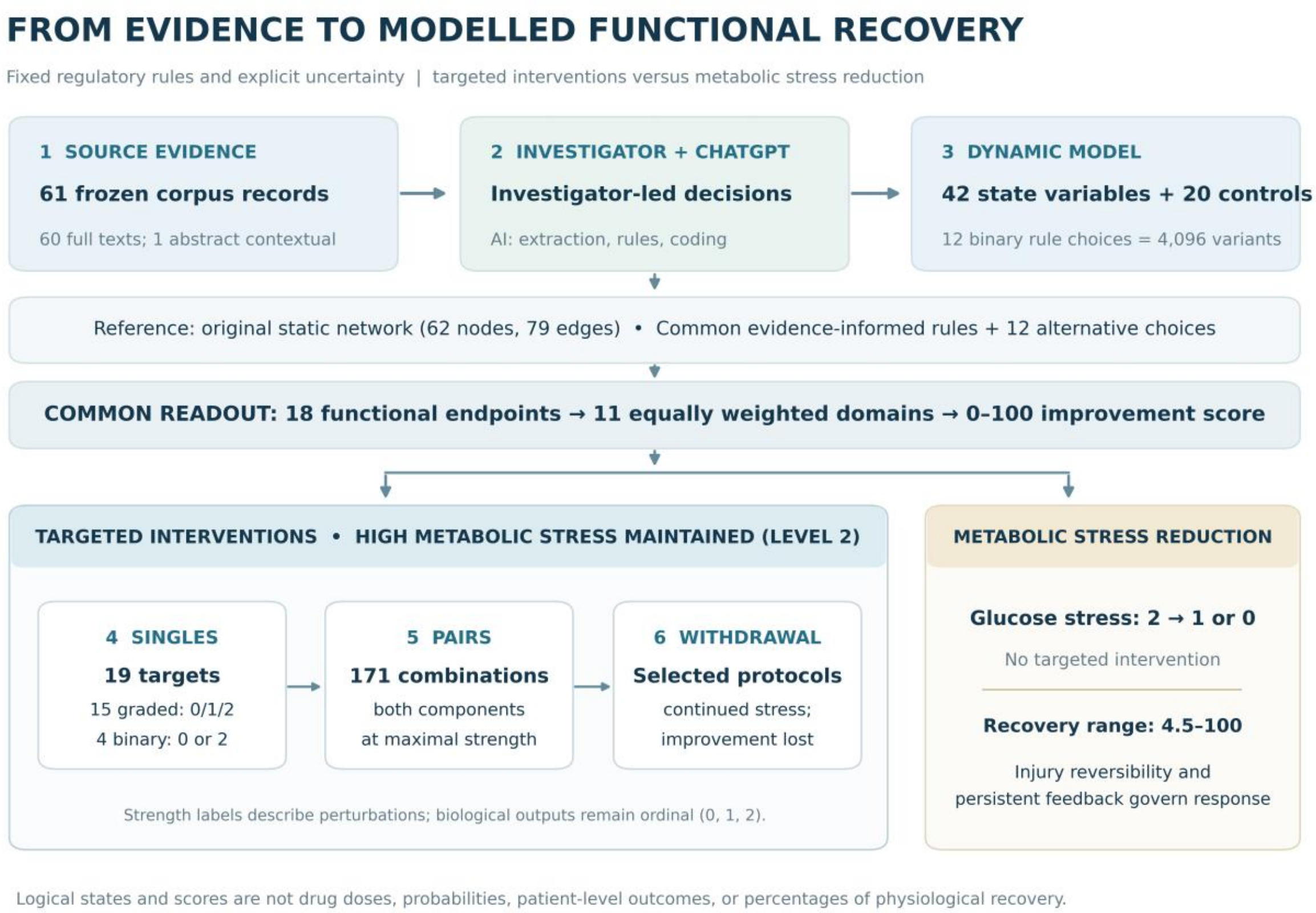

**Figure 1. Study design and workflow of the dynamic modelling analysis.** Evidence from a frozen corpus of 61 publications (60 assessed in full text and one from its abstract) informed investigator-guided model development with ChatGPT assistance. The original static network (62 nodes, 79 interactions) was preserved as a reference. The dynamic model comprised 42 functional state variables and 20 inputs or intervention controls. A common regulatory framework and 12 binary choices representing unresolved mechanisms generated 4,096 model variants. Functional improvement was quantified using 18 endpoints grouped into 11 equally weighted domains, yielding a domain-balanced score expressed on a 0–100 scale. Under persistent high metabolic stress, 19 individual interventions were assessed using graded (15 interventions; strengths 0, 1 and 2) or binary settings (four interventions; absence or maximal strength). All 171 intervention pairs were evaluated at maximal strength, followed by withdrawal simulations of selected interventions. Separately, metabolic stress reduction was examined without targeted intervention to identify regulatory conditions governing variable recovery. Ordinal states and improvement scores represent modelled responses, not calibrated drug doses, clinical probabilities or percentages of physiological recovery. AI-generated illustrative image (OpenAI image generation), reviewed by the author.

| Targeted intervention | Absent 0 | Intermediate 1 | Maximal 2 | Domains* |
|---|---|---|---|---|
| Endothelial ROS suppression | 0.0 | 12.1 | 24.2 | 3 |
| Endothelial Akt augmentation | 0.0 | 7.6–10.6 | 15.2–21.2 | 2 |
| Haematopoietic p66Shc reduction | 0.0 | 6.8 | 13.6 | 2 |
| OSM blockade | 0.0 | 6.8 | 13.6 | 2 |
| ROCK inhibition | 0.0 | 6.1 | 12.1 | 2 |
| Acid sphingomyelinase inhibition | 0.0 | – | 9.1 | 1 |
| Angiopoietin-2 blockade | 0.0 | – | 9.1 | 1 |
| CXCR4 blockade | 0.0 | – | 9.1 | 1 |
| Jmjd3 inhibition | 0.0 | – | 9.1 | 1 |
| Pericyte Akt augmentation | 0.0 | 4.5 | 9.1 | 1 |
| STAT3 blockade | 0.0 | 4.5 | 9.1 | 1 |
| Stromal p66Shc reduction | 0.0 | 4.5 | 9.1 | 1 |
| CCR2 blockade | 0.0 | 0.0 | 4.5 | 1 |
| MCP-1 suppression | 0.0 | 0.0 | 4.5 | 1 |
| NF-κB suppression | 0.0 | 0.0–9.1 | 4.5–18.2 | 1 |
| RAGE blockade | 0.0 | 0.0–9.1 | 4.5–18.2 | 1 |
| S100A8/A9 suppression | 0.0 | 0.0–9.1 | 4.5–18.2 | 1 |
| CGRP augmentation | 0.0 | 1.5 | 3.0 | 1 |
| cAMP bypass (forskolin) | 0.0 | 1.5 | 3.0 | 1 |

Scores show minimum gain across regulatory variants (out of 100).

– = not modelled at intermediate strength (four binary interventions). Hatched edge = rule-dependent range.

*Domains consistently improved at maximal strength (of 11). Strengths are not calibrated doses.

**Figure 2. Graded responses to individual interventions under persistent high metabolic stress.** All 19 targeted interventions were assessed from the same diabetic baseline across 4,096 regulatory model variants with glucose stress maintained at level 2. Comparative strength 0 denotes no intervention, 1 an intermediate perturbation where defined, and 2 the maximal predefined setting. Fifteen interventions had three settings; four (angiopoietin-2, Jmjd3, acid sphingomyelinase and CXCR4 blockade) were assessed only at strengths 0 and 2 and show an em dash at strength 1. Some of the other intermediate settings were exploratory constraints on molecular or regulatory states and were not experimentally calibrated doses. Cells give the minimum domain-balanced improvement score (0–100) or the minimum–maximum range where alternative rules produce different results; gold hatching denotes additional rule-dependent improvement. The right column shows the number of domains improved in all rule variants at maximal strength (out of 11). All maximal-strength responses reproduce the original model. The score integrates favourable and adverse ordinal changes across 18 endpoints and is not a percentage of physiological recovery. AI-generated illustrative image (OpenAI image generation), reviewed by the author.

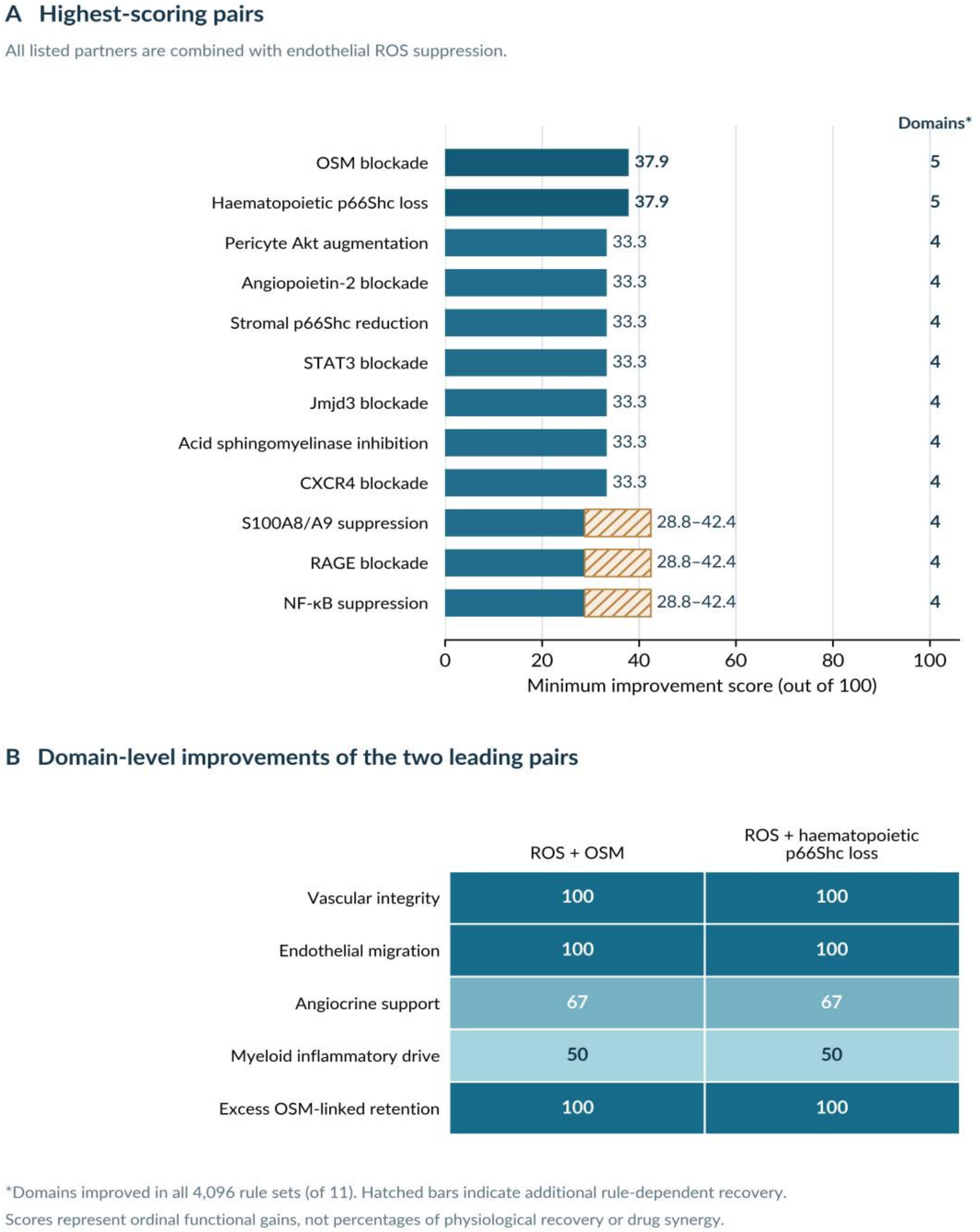

**Figure 3.**

**Highest-ranking paired interventions under persistent high metabolic stress.** (A) Twelve selected leading combinations from the complete screen of 171 unordered pairs. All displayed pairs contain endothelial reactive oxygen species (ROS) suppression; the second intervention is listed. Each component was applied at maximal strength under high metabolic stress. Bars indicate the minimum domain-balanced improvement score (out of 100) across 4,096 rule variants; hatched extensions show rule-dependent higher scores. The right column reports domains consistently improved (out of 11). Among pairs sharing the cutoff minimum score of 28.8, those with the larger maxima were selected for display; the complete ranking remains in Supplementary Section S5. (B) Minimum direction-adjusted domain improvements for the two leading pairs: ROS suppression with oncostatin M (OSM) blockade or haematopoietic p66Shc loss. The five consistently improved domains are shown; the remaining six showed no improvement. Domain values describe encoded ordinal gains rather than percentage physiological restoration. The findings indicate complementary modelled responses, not pharmacological synergy. AI-generated illustrative image (OpenAI image generation), reviewed by the author.

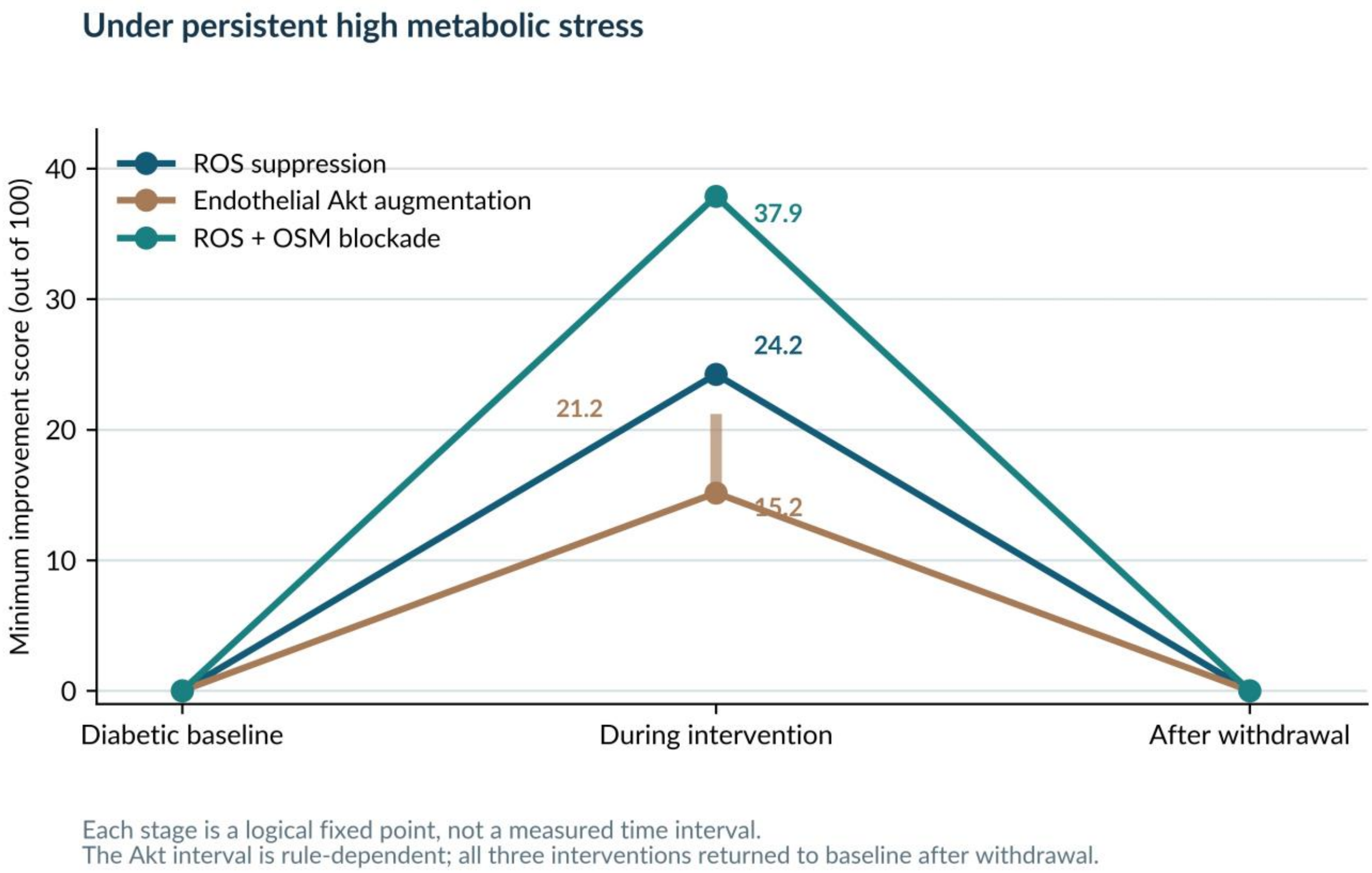

**Figure 4. Dependence of selected interventions on continued treatment under persistent high metabolic stress.** Three staged simulations compared the diabetic baseline, the model fixed point during maximal endothelial ROS suppression, maximal endothelial protein kinase B (Akt) augmentation, or combined ROS suppression–OSM blockade, and the fixed point after complete intervention withdrawal. Glucose stress remained at level 2 throughout. Plotted values are the minimum domain-balanced improvement scores across 4,096 regulatory variants. The Akt response also had a rule-dependent maximum of 21.2, indicated by the vertical interval at the active stage. Every rule variant returned to the diabetic baseline after withdrawal (score 0). These are successive logical fixed points, not calibrated exposure times or clinical durability estimates. Other interventions were not examined in this high-stress withdrawal comparison. AI-generated illustrative image (OpenAI image generation), reviewed by the author.

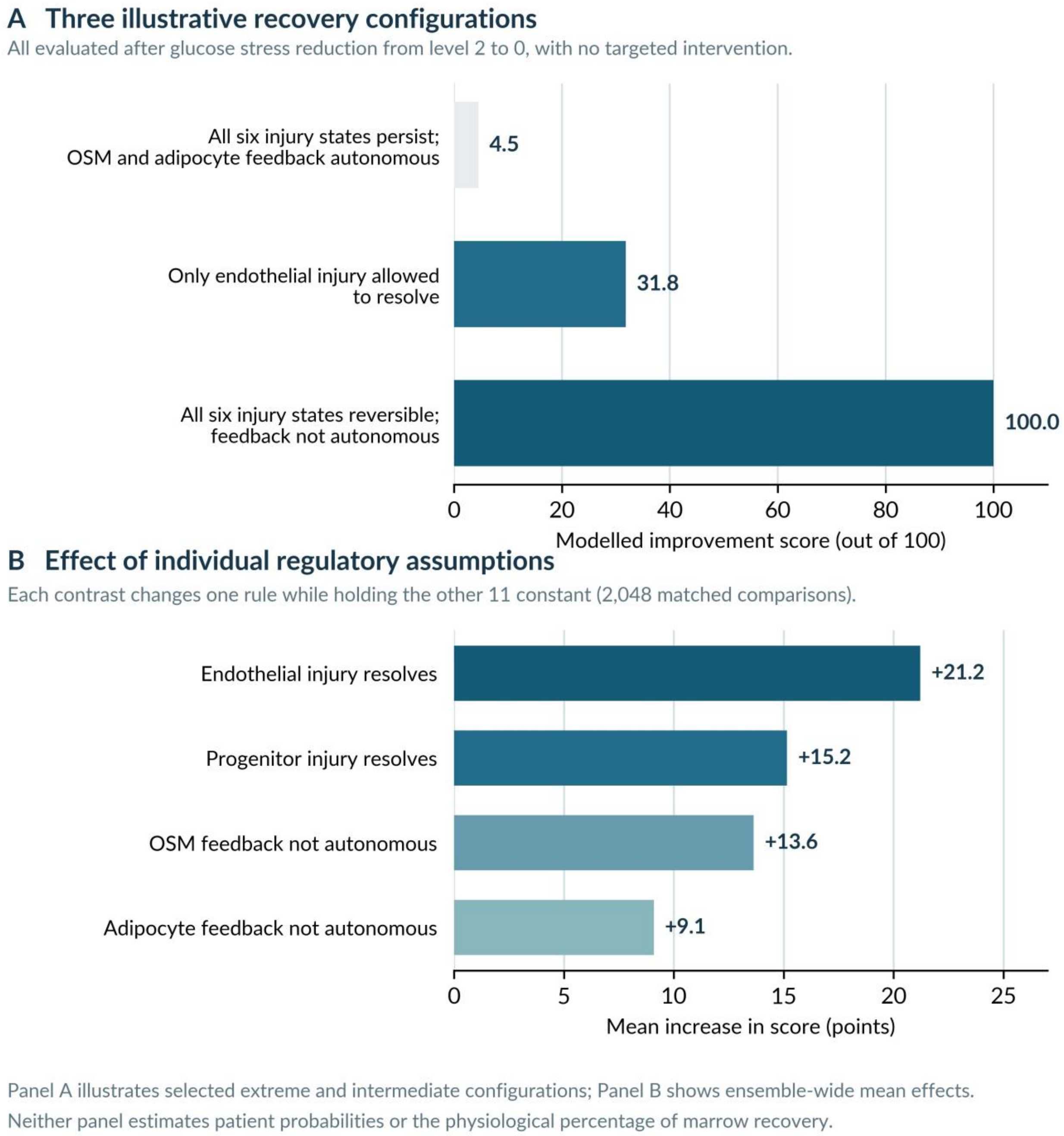

**Figure 5. Regulatory conditions determine recovery after metabolic stress reduction alone.** Metabolic stress was reduced from level 2 to 0 without a targeted intervention. (A) Three representative configurations from the 4,096 regulatory variants. Persisting dysfunction in all six injury states and autonomous OSM and adipocyte feedback resulted in an improvement score of 4.5. Permitting endothelial injury alone to resolve increased the score to 31.8. Allowing all six injury states to resolve and eliminating autonomy of both feedback circuits yielded 100, representing complete restoration only of the 18 selected scored endpoints. (B) Mean score changes when one regulatory assumption was switched across 2,048 otherwise matched rule pairs. The four most influential reversible conditions are shown. Unlike the illustrative scenarios in A, these are mean effects across matched variants. Values describe sensitivity to encoded assumptions, not causal effects measured experimentally, probabilities of recovery, or biological kinetics. AI-generated illustrative image (OpenAI image generation), reviewed by the author.

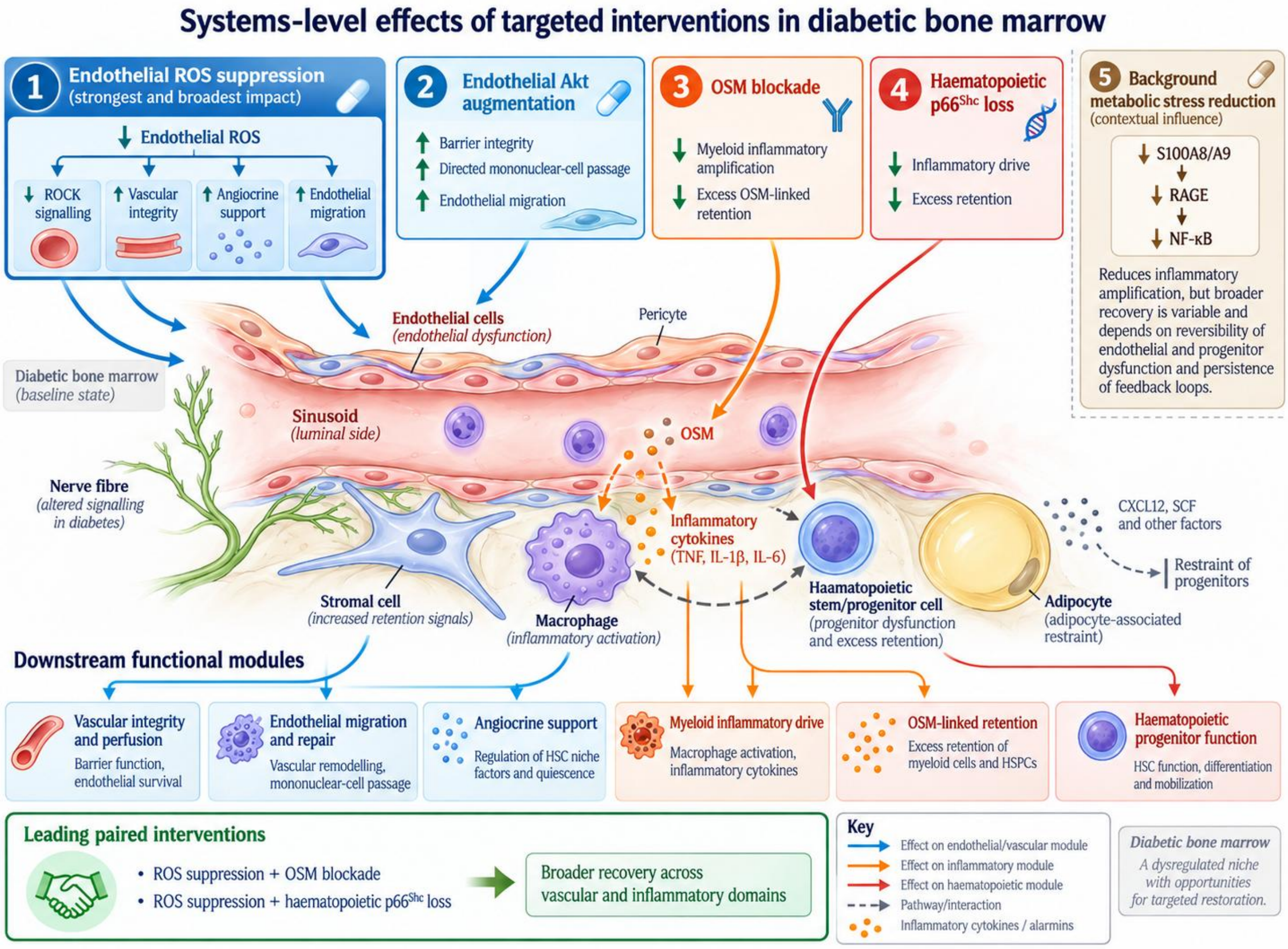

**Figure 6. Conceptual systems-level summary of targeted intervention effects in diabetic bone marrow.** Schematic representation of the modelled vascular and inflammatory pathways. Endothelial ROS suppression improves vascular integrity, migration and angiocrine support; endothelial Akt augmentation improves selected endothelial functions. OSM blockade and haematopoietic p66Shc loss reduce inflammatory drive and excess retention. Combining ROS suppression with either OSM blockade or haematopoietic p66Shc loss extends improvement across complementary functional domains. Metabolic stress reduction is represented separately as a background condition; broader recovery depends on reversibility of established dysfunction and persistence of feedback. The illustration is conceptual and does not represent measurements, treatment efficacy, or pharmacological synergy. AI-generated illustrative artwork (OpenAI image generation), reviewed by the author.